\documentclass[letterpaper,conference,final]{IEEEtran}

\usepackage[binary-units=true]{siunitx}
\DeclareSIUnit \dBm {dBm}
\DeclareSIUnit \dB {dB} 
\DeclareSIUnit \dBi {dBi} 
\DeclareSIUnit \Kbps {Kbps}
\DeclareSIUnit \Mbps {Mbps}
\DeclareSIUnit \Gbps {Gbps}
\DeclareSIUnit \kBps {kBps}
\DeclareSIUnit \MBps {MBps}
\DeclareSIUnit \GBps {GBps}

\usepackage{cite}      
\usepackage[caption=false, font=footnotesize]{subfig}
\usepackage[dvips]{graphicx}
\usepackage{url}       
\usepackage{amsfonts}
\usepackage{amssymb}
\usepackage{mathtools}
\usepackage{times}
\usepackage{algpseudocode}
\usepackage{fourier}
\usepackage{algorithm}
\usepackage{enumerate}
\usepackage{multirow}
\usepackage{tikz}
\usepackage{array}
\usepackage{gensymb}
\usepackage{eucal}
\usepackage{mathtools}
\usepackage{nicefrac}
\usepackage{tabularx}

\newcolumntype{P}[1]{>{\centering\arraybackslash}p{#1}}

\bstctlcite{IEEEexample:BSTcontrol}

\algnewcommand\algorithmicinput{\textbf{Level 1:}}
\algnewcommand\Level{\item[\algorithmicinput]}

\algnewcommand\algorithmicinputt{\textbf{Level 2:}}
\algnewcommand\Levell{\item[\algorithmicinputt]}

\algnewcommand\algorithmicinputtt{\textbf{Level 3:}}
\algnewcommand\Levelll{\item[\algorithmicinputtt]}

\algnewcommand\algorithmicinputttt{\textbf{Output:}}
\algnewcommand\Output{\item[\algorithmicinputttt]}

\usepackage[english]{babel}
\usepackage{blindtext}
\IEEEoverridecommandlockouts
\begin{document}
\title{On Urban Traffic Flow Benefits of\\ Connected and Automated Vehicles}

\author{\IEEEauthorblockN{Ioannis Mavromatis\IEEEauthorrefmark{1}\IEEEauthorrefmark{2}, Andrea Tassi\IEEEauthorrefmark{2}\thanks{The first and second authors equally contributed to the paper.}, Robert J. Piechocki\IEEEauthorrefmark{2}\IEEEauthorrefmark{3}, and Mahesh Sooriyabandara\IEEEauthorrefmark{1}}
\IEEEauthorblockA{\IEEEauthorrefmark{1} Toshiba Research EU Ltd.
Bristol, UK\\
\IEEEauthorrefmark{2} Department of Electrical and Electronic Engineering, University of Bristol, UK\\
\IEEEauthorrefmark{3}The Alan Turing Institute, London, NW1 2DB, UK\\
Email: Ioannis.Mavromatis@toshiba-trel.com, \{A.Tassi, R.J.Piechocki\}@bristol.ac.uk, Mahesh@toshiba-trel.com}\vspace*{-10mm}}

\maketitle

\begin{abstract}
Automated Vehicles are an integral part of Intelligent Transportation Systems (ITSs) and are expected to play a crucial role in the future mobility services. This paper investigates two classes of self-driving vehicles: (i) Level 4\&5 Automated Vehicles (AVs) that rely solely on their on-board sensors for environmental perception tasks, and (ii) Connected and Automated Vehicles (CAVs), leveraging connectivity to further enhance perception via driving intention and sensor information sharing. Our investigation considers and quantifies the impact of each vehicle group in large urban road networks in Europe and in the USA. The key performance metrics are the traffic congestion, average speed and average trip time. Specifically, the numerical studies show that the traffic congestion can be reduced by up to a factor of four, while the average flow speeds of CAV group remains closer to the speed limits and can be up to $300\%$ greater than the human-driven vehicles. Finally, traffic situations are also studied, indicating that even a small market penetration of CAVs will have a substantial net positive effect on the traffic flows. 
\end{abstract}

\vspace{-3mm}
\section{Introduction}

By 2025, autonomous vehicles requiring little or no interaction with a human driver (Autonomy Level 4 and 5, respectively) are expected to hit the roads worldwide~\cite{MRQ}. Both the European Commission's C-ITS initiative and the U.S. Department of Transportation estimate that automated driving will contribute to reducing traffic jams, take advantage of the full capacity of road networks, and improve the safety of vulnerable road users~\cite{C_ITS,NHTSA}.

Automated vehicles, equipped with sensors and Vehicle-to-Everything (V2X) communication interfaces, will share their collected information and driving intentions with the surrounding environment. Access to such information can enable a cooperative element among the vehicles, enhance the traffic efficiency by making faster decisions, and adapt to the different traffic conditions~\cite{cooperativeDriving}. Based on their communication capabilities, vehicles can be classified either as: (i) Autonomous Vehicles (AV), providing autonomous features solely based on the information collected by the vehicle itself, (ii) Connected and Autonomous Vehicles (CAVs), capable of both collecting and sharing this information with the surrounding environment.

In this paper, we investigate the impact of both AVs and CAVs on large urban road networks. Our investigation focuses on the traffic flow, the network capacity, the average speed, and trip time achieved under different traffic load conditions. As our baseline measurement, we consider traditional Human-Driven (HD) vehicles.

The benefits of AVs and CAVs have been demonstrated in the past. For example, authors in~\cite{trafficImplications} found that a purely CAV scenario could improve the highway capacity by $300\%$, while purely AV scenarios show relatively indifferent performance. When AVs are assigned a dedicated lane on a highway, the road throughput is enhanced significantly, as shown in~\cite{effectReservedLanes}. These works, even though they reflect the benefits of AVs and CAVs, they do not consider urban traffic. In our approach, we will focus on the more demanding urban scenarios. This was the case in~\cite{effectAutonomousVehicles}, where a $40\%$ increase was observed in the city traffic flow, as perceived at a single-intersection scenario. Similarly, in~\cite{impactUrbanNetwork}, the authors observed a $16\%$ improvement in the maximum urban traffic flow under an artificially generated grid-like network, considering a $100\%$ AV adoption rate, and a $25\%$ benefit on a small-scale real-world map. However, both works did not consider the differences that may arise from the unique layouts of the different real-world cities. Furthermore, neither the cooperation between the vehicles nor the different congestion levels were taken into account. To the best of our knowledge, no work in the literature shows the benefits of both AVs and CAVs under large-scale urban environments and different city layouts.

With these regards, this paper aims to fill this gap, providing an extensive performance investigation of fleets of AVs and CAVs in large and realistic road networks. In particular, our investigation assesses the impact of the unique vehicle characteristics on the different road layouts and traffic congestion levels over five European and American cities and $\SI{165}{\kilo\meter}$ of total road length per-road network. Finally, the impact of the lack of connectivity on CAV-based scenarios is also considered, showing the importance of V2X communication links as the traffic demand is increased. 

The rest of this paper is organized as follows. Sec.~\ref{sec:systemDescription} presents the considered experimental setup and assumptions around the dynamics of HD vehicles, AVs, and CAVs. Sec.~\ref{sec.III} describes the processing of the cartographic data and how the mobility traces were generated for all vehicles. The simulated scenarios and our findings are discussed in Sec.~\ref{sec:results}. Finally, in Sec.~\ref{sec:conclusions}, we draw our conclusions.

\section{System Description}\label{sec:systemDescription}

Consider an urban area consisting of several buildings, areas of foliage and vegetation, and an extensive road network. A vehicle on the road coexists with its surrounding vehicles, pedestrians, cyclist, and all means of public transport. A human driver usually observes the surrounding environment and reacts accordingly to any changes (for e.g., a traffic light turns red, or reaches a junction and has to yield the right-of-way). On the other hand, an AV relies on its on-board sensors to react to the same changes. Finally, a CAV exchanging sensor data and driving intentions is capable of artificially generating a ``birds-eye'' view of the surrounding environment, perform smoother cooperative maneuvers on the road, and plan its long-term route in a more sophisticated fashion.  

The reaction time of AVs is considerably shorter when compared to humans. Studies on existing Autonomy Level 2 and 3 vehicles~\cite{reactionTime}, showed that AVs can react on average in \textasciitilde\SI{0.66}{\second} while humans require over $\SI{0.9}{\second}$ or even \SI{1.5}{\second}, in impaired scenarios. Moving towards full autonomy and CAVs, these numbers are expected to decrease even further.

When studying the impact of AVs and CAVs, all the above interactions and distinct characteristics should be taken into account. For our evaluation, we utilized the Simulation of Urban Mobility (SUMO) traffic generator~\cite{sumo}. SUMO is capable to microscopically model and control each vehicle and traffic light and provides a flexible framework for generating finely-tailored traffic management scenarios. 

A developed MATLAB framework accompanies our investigation. Our framework is publicly available under {\tt\small https://github.com/v2x-dev/sumoCAVs} repository. It can be used to automate the simulation steps, enables the easier generation and validation of different scenarios and works in two phases. At first, we generate the required structures and files for SUMO, based on several user-defined parameters. Later, using TraCI4Matlab framework~\cite{traci4matlab}, the different scenarios are executed in a parallelized fashion.
The generated data are logged as MAT-files and can be processed by the provided scripts. More details about the provided capability can be found in the next sections.

\subsection{The Car-Following Model}\label{sub:carFollowing}

In the field of transportation engineering, a traffic flow is defined as the study of the interactions between travelers (for e.g., vehicles, pedestrians, cyclists, etc.) and the road infrastructure network (e.g., traffic lights, signage, etc.). A car-following model describes how vehicles follow one another and interact on a roadway. In this paper, we refer to the Intelligent Driver Model (IDM)~\cite{idmModel} as our chosen car-following model, for all the different vehicle types (HD, AV, CAV). 

Comparing IDM to other car-following models (for e.g. Gipp's~\cite{gippsModel} and Krauss~\cite{kraussModel} models), it better represents the macroscopic reality. Authors in~\cite{comparisonModelsMesoscopic} showed that under a steady-state (using a calibrated model for each scenario), IDM achieves lower traffic speed errors and better describes the road capacity and vehicle density. Microscopically, IDM still outperforms the above-mentioned models. As reported in~\cite{comparisonModelsMicroscopic}, IDM generates more realistic vehicle trajectories when crossing an intersection. Also, IDM is capable of simulating flow conserving inhomogeneities (for e.g., on-/off-highway ramps, lane closing, etc.), the coexistence of moving localized vehicles clusters and clusters pinned at road inhomogeneities, as well as congested traffic behaviors, all use-cases that can be concurrently observed on large-scale city-wide scenarios~\cite{idmModel}. For the above reasons, IDM was chosen as the most suitable model for our study. 

\subsection{Uniform Scenario Generation for all Cities}
As discussed in~\cite{realisticSUMO}, to achieve a true-to-life representation of the real-world when generating a scenario in SUMO, different variables should be taken into consideration. These are the demographic information and buildings from specific areas, information about the multimodal public transport system for each city, parking spaces, available pedestrianized streets or cycling paths, etc. However, given the large-scale nature of our investigation, many of this information is not available for some or all our cities-of-interest. Also, the above information is city-specific and varies between the scenarios. 

Introducing these variables increases the complexity of isolating the effect of the city layout on the different traffic flows. Therefore, for our investigation, we consider a uniform way for generating the different scenarios, still introducing real-world behaviors (for e.g., choosing left- or right-hand traffic). What is more, considering different vehicle densities in each city, we can simulate different traffic patterns, for example, low traffic conditions (during the evening hours) or heavy traffic conditions (rush hours). In the next section, we describe in more detail the way that each scenario was generated.

\section{Scenario Generation and Execution}\label{sec.III}

A SUMO scenario requires a set of files describing the scenario parameters. Investigating a scenario requires several steps. Our framework is capable of automating the process, starting from Step 2. Briefly, the steps are as follows: (i) Download a map from OpenStreetMap (OSM)~\cite{OpenStreetMap}, (ii) Convert the map into a SUMO network file (this is described as a road network in SUMO), (iii) Based on the user-defined vehicle characteristics, generate the different vehicle type files for each scenario, (iv) Generate the mobility traces, (v) Based on the road network and mobility traces, generate the traffic light adaptation and coordination files, (vi) The scenario is executed, and the data are saved in a MATLAB-file format. In the post-processing phase, the user can choose to generate all the results. The following sections describe each step in more detail.

\subsection{Map Conversion and Traffic Light Generation}\label{sub:mapTrafficLights}

Our study starts by choosing the appropriate maps from OSM~\cite{OpenStreetMap} and importing them into SUMO. An OSM map (\textit{.osm} file) stores in an XML-formatted file all the cartographic details of an area such as the road types (e.g., bus lanes, regular roads, pathways, etc.), buildings, foliage, traffic lights, etc. Even though the chosen cities are well-mapped in OSM, there are several flaws that should be considered when converting a map with \textit{netconvert} tool~\cite{sumo}.

At first, many streets (called \textit{edges} in SUMO) are considered as legitimate U-turns, a behavior not common in the real-world. Disabling that feature (\textit{-{}-no-turnarounds}), enables a more normal driving behavior and promotes through-traffic. Furthermore, \textit{-{}-roundabouts.guess} was used to obtain the correct priorities for all roundabouts. Most imported OSM maps do not provide information about highway on- and off-ramps. To correct this behavior, the ramp-guessing was enabled using \textit{-{}-ramps.guess}. Overtaking, that is by default disabled in SUMO, was introduced using \textit{-{}-opposites.guess} and \textit{-{}-opposites.guess.fix-lengths} attributes.

OSM maps usually contain information about the traffic lights positions. However, these details could either be missing or be inaccurate. To introduce a unified and solid traffic light system, we start by guessing the positions of the missing traffic lights (using \textit{-{}-tls.guess}). The coordination between the traffic lights in close proximity ($<\SI{20}{\meter}$) is enhanced by clustering them (\textit{-{}-tls.join} and \textit{-{}-tls.guess-signals}). Finally, oddly misplaced traffic lights are discarded using  \textit{-{}-tls.discard-simple} and actuated traffic light technology is introduced, using \textit{-{}-tls.default-type actuated} that adapts the cycle phases based on the traffic demands

Finally, nowadays, humans and vehicles communicate either using traffic lights or with unspoken interactions. AVs and CAVs will rely on their on-board sensors and knowledge received from the surrounding environment to detect and predict pedestrian behavior. SUMO is not capable of providing such interaction. Therefore, only the roads that allow vehicle traffic was chosen for our investigation (using \textit{-{}-keep-edges.by-vclass}), and pedestrians were not considered.

\subsection{Generation of the Vehicle Type Distributions}\label{sub:vehicleTypes}

Based on a given network file, we can generate the mobility traces for all vehicles types, containing all the parameters that describe the vehicles behavior. The chosen parameters for our experimentation can be found in Table~\ref{tab:vehicleTypeParameters}. As described in Sec.~\ref{sub:carFollowing}, IDM was chosen as the car-following model (named as \textit{carFollowModel} in SUMO) for all the vehicle types and needs to be fine-tuned for the different vehicle types (either human or computer-driven).

AVs and especially CAVs, are expected to drive in closer proximity, at faster-pace, and provide increased safety to the passengers~\cite{speedRelaxation}. Of course, the ``safe'' margins are scenario dependent~\cite{dynamicSpeedLimits}, and can vary a lot based on parameters such as the weather, traffic jams, the quality of the road network, etc. When planning a city nowadays, the chosen fixed-speed limits represent the appropriate speed for average conditions. However, it is already a common practice to apply dynamic speed limit policies when the above parameters are changed~\cite{dynamicSpeedLimits}. For our experimentation, in order to achieve a homogeneous comparison between the different road networks, we assume that all the scenarios are under ideal conditions. This behavior is simulated using a number of parameters. At first, $\tau$ describes the drivers desired time-headway. Later, the minimum gap between two vehicles is described by \textit{minGap}. The speed factor, i.e., \textit{speedFactor}, is the multiplier of the speed limit on a road. Decreasing $\tau$ and \textit{minGap}, while increasing the \textit{speedFactor}, we can simulate different types of autonomous vehicles. The erratic behavior of human drivers of blocking a junction and create a traffic jam can be modeled using \textit{jmIgnoreKeepClearTime}. Finally, the length, the acceleration, deceleration, and acceleration exponent should be configured for each scenario. More details about the chosen parameters can be found in SUMO documentation~\cite{sumo} and our MATLAB framework.


\begin{table}[t] 
\renewcommand{\arraystretch}{1.1}
\centering
    \caption{Vehicle Type and Mobility Trace Parameters.}
    \begin{tabular}{r||c|c|c}

    \textbf{Parameter}    & \textbf{HD} & \textbf{AV}   & \textbf{CAV} \\ \hline \hline
    Desired time headway ($\tau$)~\cite{studyBehavior}              & \SI{1.69}{\second}       & \SI{0.5}{\second}         & \SI{0.1}{\second} \\
    Minimum Gap                & \SI{2.5}{\meter}       & \SI{1.5}{\meter} & \SI{1}{\meter} \\ 
    Accumulated Waiting Time~\cite{speedRelaxation} & \SI{300}{\second}     & \multicolumn{2}{c}{Disabled ($-1$)} \\
    Length                & \multicolumn{3}{c}{$ \mathcal{N}(\mu = 4.5,\, \sigma = 0.2)$} \\    
    Car-Following Model        & \multicolumn{3}{c}{Intelligent Driver Model (IDM)} \\     
    Acceleration~\cite{studyBehavior}                 & \multicolumn{3}{c}{\SI{1.25}{\meter\per\second\squared}} \\  
    Deceleration~\cite{studyBehavior}                 & \multicolumn{3}{c}{\SI{2.09}{\meter\per\second\squared}} \\        
    Acceleration Exponent~\cite{speedRelaxation}                 & \multicolumn{3}{c}{4} \\  
    Departure/Arrival Position                 & \multicolumn{3}{c}{Randomly Chosen} \\ 
    Departure/Arrival Lane                 & \multicolumn{3}{c}{Best lane/Current Lane} \\ 
    Departure/Arrival Speed & \multicolumn{3}{c}{Max Allowed/Fastest possible} \\ 
    \end{tabular}
\label{tab:vehicleTypeParameters}
\end{table}



\subsection{Mobility Trace Generation and Traffic Light Adaptation}\label{sub:mobilityTracesTrafficLights}

A typical trace file contains an entry per generated vehicle and describes several attributes of the trip, these being: (i) the vehicle type, (ii) the departure time, speed, edge and lane, (iii) the arrival position/edge, and (iv) all the intermediate edges that a vehicle will follow during the trip. Based on the above-mentioned distribution and network files, and using \textit{randomTrips} SUMO tool~\cite{sumo}, we can generate the mobility trace file for a given scenario.

Vehicles are periodically added in the simulation queue. The initial position of a vehicle is chosen at random, from all the available free positions on the plane (\textit{departPos = random\_free}). All edges on the map have an equal weight. By that, we ensure that the vehicles are uniformly spread around the city. The best available lane is always chosen, i.e., \textit{departLane = best}, allowing the vehicle to drive the longest without the need for changing lanes. Finally, the insertion speed, i.e., \textit{departSpeed = desired}, is the maximum allowed velocity, given by the speed limit multiplied by the speed factor. That ensures a continues flow of vehicles and no disruptions in the existing traffic. A random position is chosen at the departure as the arrival position, i.e., \textit{arrivalPos = random}. A maximum number of vehicles can be configured or left uncapped. 
To ensure that a desired number of vehicles always exists, we forced our generated traces to be longer than the simulation time. All these attributes are summarized in Table~\ref{tab:vehicleTypeParameters}.

Knowing the traffic flows, the traffic light cycles can be adapted to minimize the waiting time at a ``red'' light. To do that, we use \textit{tlsCoordinator} to modify the traffic-light offsets in a coordinated fashion and \textit{tlsCycleAdaptation} to modify the duration of the green phases according to Webster's formula, and achieve an actuated-style traffic light system. By that, the realism of the simulation is increased. More information about these tools can be found in SUMO documentation page~\cite{sumo}.


\subsection{Wireless Connectivity and Vehicle Rerouting}

CAVs, compared to the AVs and HD vehicles, are  capable of exchanging information with the surrounding environment. Based on that, a CAV can cooperate with the surrounding vehicles to optimize the traffic flows, and decrease the overall trip duration and length. To simulate this behavior, each CAV is equipped with a ``rerouting'' device based on a probability distribution. This device re-computes the CAV route periodically, taking into account the current and recent state of the traffic, helping CAVs to adapt their route to the traffic jams.



The rerouting of a CAV is fine-tuned based on a number of adaptation steps (past simulation steps to be considered) and an adaptation interval, i.e., the time interval between two consecutive adaptations.
Finally, the rerouting period, determines how fast vehicles can react to traffic fluctuations. If a value equals or is smaller than the adaptation interval used, then the reaction of the vehicle is almost instantaneous. These values can be found in Table~\ref{tab:rerouting}.

\begin{table}[t] 
\renewcommand{\arraystretch}{1.1}
\centering
    \caption{Rerouting Parameters for CAVs.}
    \begin{tabular}{r||c|c|c}
    \textbf{Parameter}    & \textbf{HD} & \textbf{AV}   & \textbf{CAV} \\ \hline \hline
    Rerouting Device & \multicolumn{2}{c}{Not Available} & Available \\ 
    Adaptation Steps & \multicolumn{2}{c}{-} & \SI{60}{} \\ 
    Adaptation Interval & \multicolumn{2}{c}{-} & \SI{1}{\second} \\ 
    Rerouting Period & \multicolumn{2}{c}{-} & \SI{0}{\second} \\ 
    \end{tabular}
\label{tab:rerouting}
\end{table}

\subsection{Execution of the Scenarios}\label{sub:simulationExecution}
SUMO configuration file (called \textit{sumocfg}) links everything described above and contains several simulation parameters. Our MATLAB framework, bidirectionally interacts with SUMO using TraCI4Matlab framework~\cite{traci4matlab}, dynamically generates the different required files and executes the SUMO tool commands. Finally, it saves the different states of the simulation (for e.g., the vehicle positions per time step) and overall statistics at the end.


The execution of the scenarios needs to be fine-tuned. At first, the teleportation of vehicles is disabled, setting \textit{time-to-teleport} greater than the execution time and disabling \textit{collision.action}. By that, we avoid vehicles disappearing under heavy traffic jam conditions or when they collide (this being the default SUMO behavior). Also, within a simulation environment, the real reaction time is fundamentally limited by the simulation step, i.e., \textit{step-length}. Therefore, $\tau$, described in Sec.~\ref{sub:vehicleTypes}, should always be greater than the step duration. Finally, in order to guarantee that all the routes generated are up-to-date, we ensure that vehicles are either added in the simulation on their designated departure time or are being discarded from the simulation queue. We do that by setting \textit{max-depart-delay} equal to the simulation step. The values for all these parameters are summarized in Table~\ref{tab:sumoParameters}.

\subsection{Fluid Bounding Boxes}\label{sub:fluidBox}

Given the unique city layouts and road networks, we consider a fluid bounding box that restricts the area-of-interest for some Key Performance Indicators (KPIs). The reason behind that is two-fold. At first, we have the abnormal behavior when a vehicle approaching the map boundaries. When an OSM map is converted to a SUMO network file, the connections of the roads close to the edge of the map might be corrupted. Introducing the bounding box, we can exclude these results from our evaluation. Furthermore, for KPIs such as the vehicle density, when investigated and compared between different scenarios and different city layouts, it is  crucial to refer to a similar road length~\cite{fluidBoundingBox}. Therefore, a suitable metric related to the size of the map, accounting for the total length of the roads present in a given area was considered. 

For our investigation, we introduced the \textit{overall target distance} $D$ of the road network. $D$ takes into account all the above and is calculated as follows:
\begin{equation}
D = \frac{\sum_{s = 1}^{S_a} d_s \* l_s}{a}
\end{equation}
where $d_s$ is the length of a street $s$ (measured in meters), and $l_s$ is the number of lanes of that street. Also, $S_a$ is the set of all streets within the area $a$ (measured in $\SI{}{\kilo\meter}^2$). Furthermore, we introduce the \textit{map-specific distance} $D^{\prime}$. Starting from the center of the map, we iteratively choose the next closest road to the map center and add it to $D^{\prime}$. The iteration stops when $D^{\prime}>D$, and we remove the last added edge from $D^{\prime}$. Having $D^{\prime}$, and taking the  furthest top-right and bottom-left coordinates of all the edges and we have our map-specific bounding box and ensure that roughly the same road length is enclosed for all the different scenarios.


\begin{table}[t] 
\renewcommand{\arraystretch}{1.1}
\centering
    \caption{Scenario Configuration Parameters.}
    \begin{tabular}{r||c|c}
    \textbf{Parameter} & Uncapped & Capped \\ \hline \hline
    Max. No. of Vehicles & Uncapped & $\left\{250:250:1500\right\}$ \\ 
    Insertion Rate ($r$) & $\left\{0.8:0.05:1\right\}$\SI{}{\second} & \SI{0.1}{\second} \\
    Time to teleport & \multicolumn{2}{c}{\SI{3601}{\second}} \\ 
    Collision action & \multicolumn{2}{c}{None} \\ 
    Simulation Time & \multicolumn{2}{c}{\SI{3600}{\second}} \\ 
    Timestep Length & \multicolumn{2}{c}{\SI{100}{\milli\second}} \\
    Max. Departure Delay &\multicolumn{2}{c}{\SI{100}{\milli\second}} \\
    Target Distance $D$ & \multicolumn{2}{c}{$\SI{130}{\kilo\meter}$} \\
    \end{tabular}
\label{tab:sumoParameters}
\end{table}

\begin{table*}[t] 
\renewcommand{\arraystretch}{1.1}
\centering
    \caption{List of Map Areas Used and their Road Network Characteristics.}
    \begin{tabular}{r||c|c|c|c|c}
    \textbf{Parameter}    & \textbf{Manhattan, USA} & \textbf{Paris, FR}   & \textbf{Berlin, DE} & \textbf{Rome, IT} & \textbf{London, GB} \\ \hline \hline
    Centre & $73.98585\degree W, 40.75495\degree N$   &  $2.3004\degree E, 48.875\degree N$  & $13.3969\degree E, 52.53645\degree N$  & $12.5143\degree E, 41.8823\degree N$ & $0.1638\degree W, 51.5207\degree N$ \\ 
    Total Road length & \SI{168.68}{\kilo\meter} & \SI{160.39}{\kilo\meter} & \SI{161.86}{\kilo\meter} & \SI{167.70}{\kilo\meter} & \SI{168.48}{\kilo\meter}  \\ 
    Map Size $\left[ \mathcal{M}_{x}, \mathcal{M}_{y} \right]$ & \SI{3.776}{\kilo\meter} $\times$ \SI{3.264}{\kilo\meter} & \SI{3.343}{\kilo\meter} $\times$ \SI{2.632}{\kilo\meter} & \SI{3.353}{\kilo\meter} $\times$ \SI{2.955}{\kilo\meter} & \SI{2.985}{\kilo\meter} $\times$ \SI{3.163}{\kilo\meter} & \SI{2.564}{\kilo\meter} $\times$ \SI{2.373}{\kilo\meter} \\
    Traffic Lights & 292 & 75 & 19 & 22 & 62 \\
    Total Junctions & 1939 &  2933 & 3339 & 3684 & 5879 \\
    Priority Junctions & 594 &  1134 & 20 & 1597 & 2283 \\
    Roundabouts & 0 &  8 & 1 & 6 & 0 \\
    \end{tabular}
\label{tab:maps}
\end{table*}

\section{Performance Evaluation}\label{sec:results}

To investigate the effect of the road layout on the traffic flows, we chose five different cities, i.e., Manhattan, Paris, Berlin, Rome, and London.  In particular, Manhattan has a grid-like road network regulated with a large number of traffic lights. Paris and London, follow a loose- and uneven-grid-like architecture, with a combination of main arteries and side roads. However, London consists of a vast number of roads and junctions, while Paris relies on roundabouts to connect its main arteries. Berlin, on the other hand, has a spiderweb-like layout with a small number of priority roads. Finally, the unique characteristic of Rome is the very narrow roads and increased number of non-vehicle designated areas that limit the routing options for a driver. For each city, a map with a road length of \textasciitilde\SI{165}{\kilo\meter} was chosen to provide a similar simulation environment across the different scenarios. Table~\ref{tab:maps} summarizes the static network information for all maps.

To understand the unique characteristics of the different layouts and the improvement that AVs and CAVs may bring, we focused on two different use-cases. The first one is an uncapped scenario, where the number of vehicles per time slot is not restricted. Five different insertion rates $r$ were considered that regulate the number of vehicles inserted in the simulation time -- this means that SUMO adds a new vehicle into the simulation every $r$ second. On the other hand, we have capped scenarios. For that, the number of vehicles is restricted by a maximum threshold value. When this number is reached, a new vehicle is added in the simulation only when one of the existing vehicles arrives at its destination and is removed. The traffic flows start from light traffic (\SI{250}{} vehicles) and go up to a severe traffic jam (\SI{1500} vehicles). All the parameters for these two use-cases can be found in Table~\ref{tab:sumoParameters}. As discussed in Sec.~\ref{sub:mobilityTracesTrafficLights}, the actuated traffic lights can enhance the traffic flows in the city. Therefore, for all the scenarios, actuated traffic lights were considered.

\begin{figure}[t]     
\centering
\includegraphics[width=1\columnwidth]{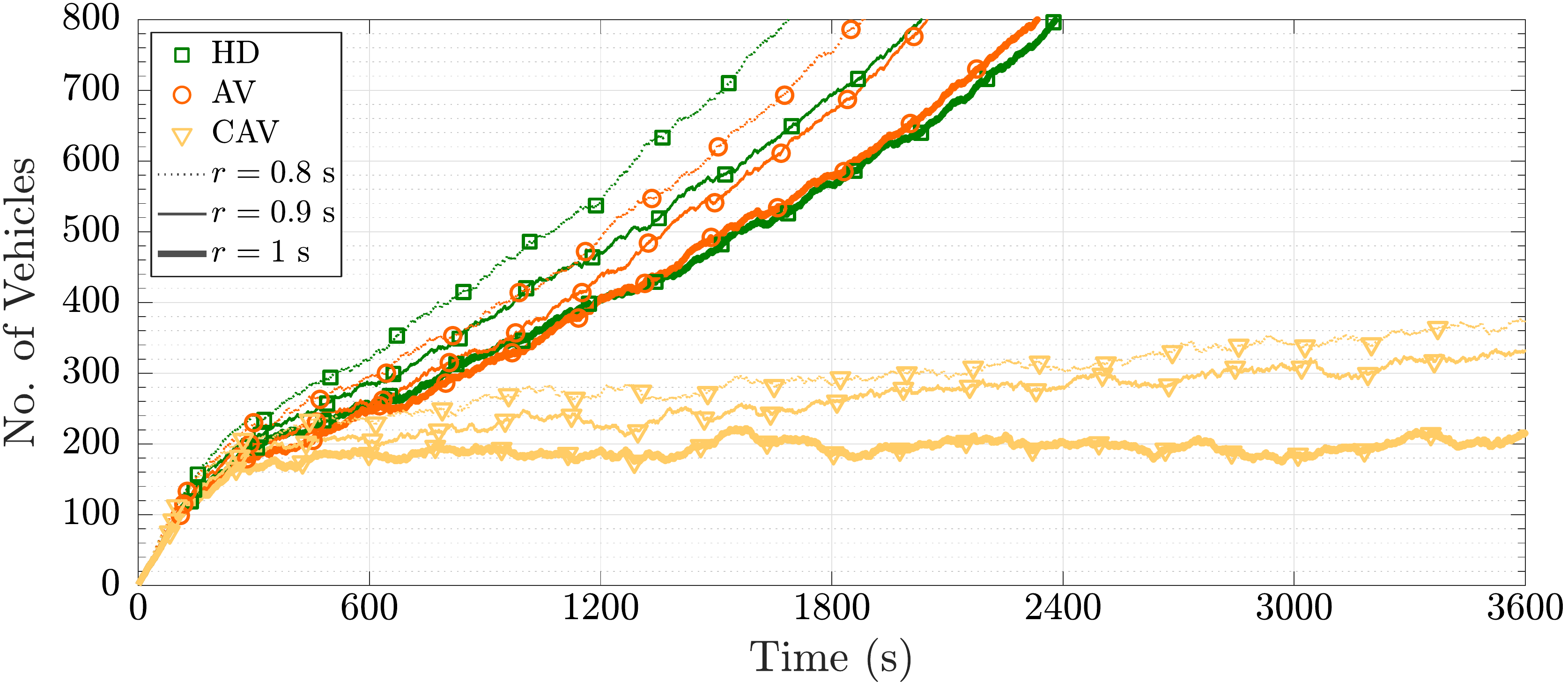}
    \vspace{-5.5mm}
    \caption{Number of vehicles for the uncapped scenario for Paris, as a function of time and different insertion rates.}
    \label{fig:uncappedParis}
\end{figure}

\begin{figure}[t]     
\centering
\includegraphics[width=1\columnwidth]{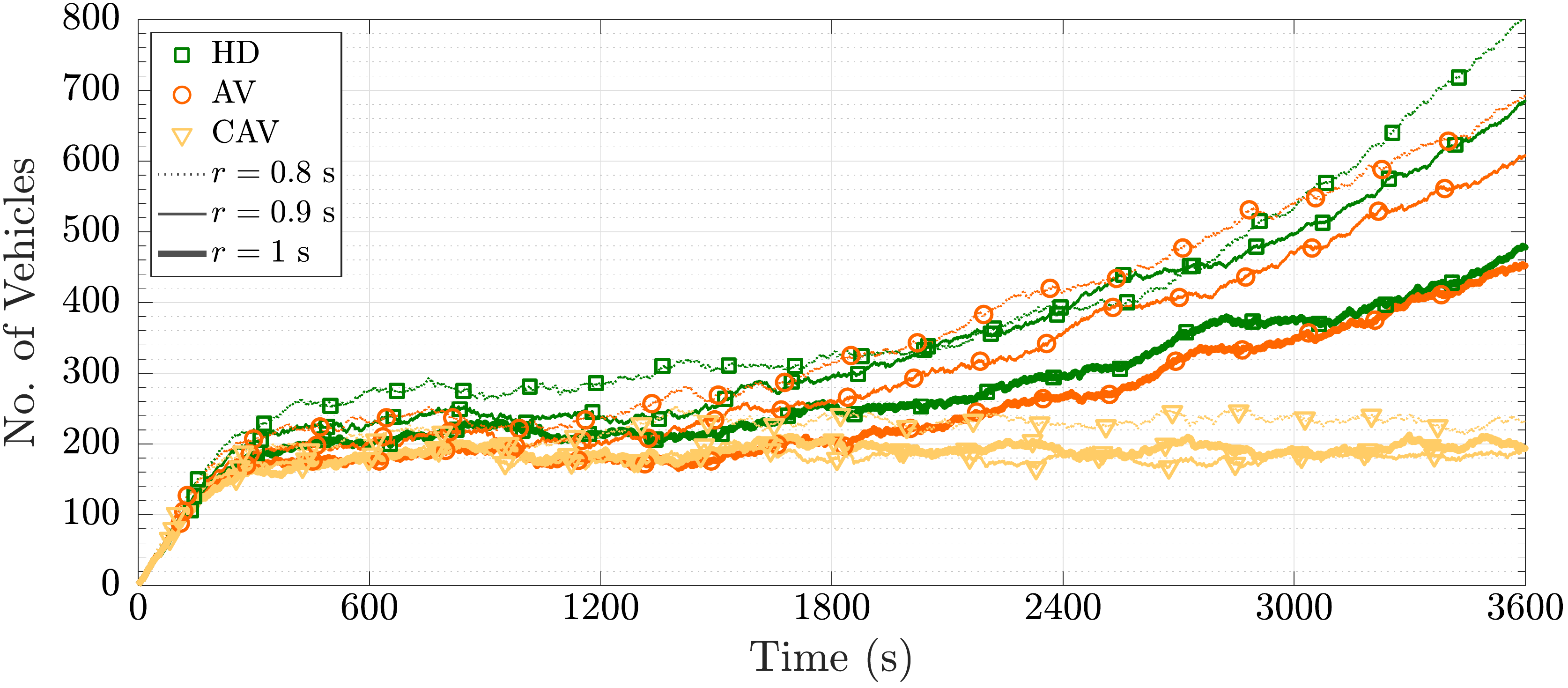}
    \vspace{-5.5mm}
    \caption{Number of vehicles for the uncapped scenario for London, as a function of time and different insertion rates.}
    \label{fig:uncappedLondon}
\end{figure}

\begin{figure}[t]     
\centering
\includegraphics[width=1\columnwidth]{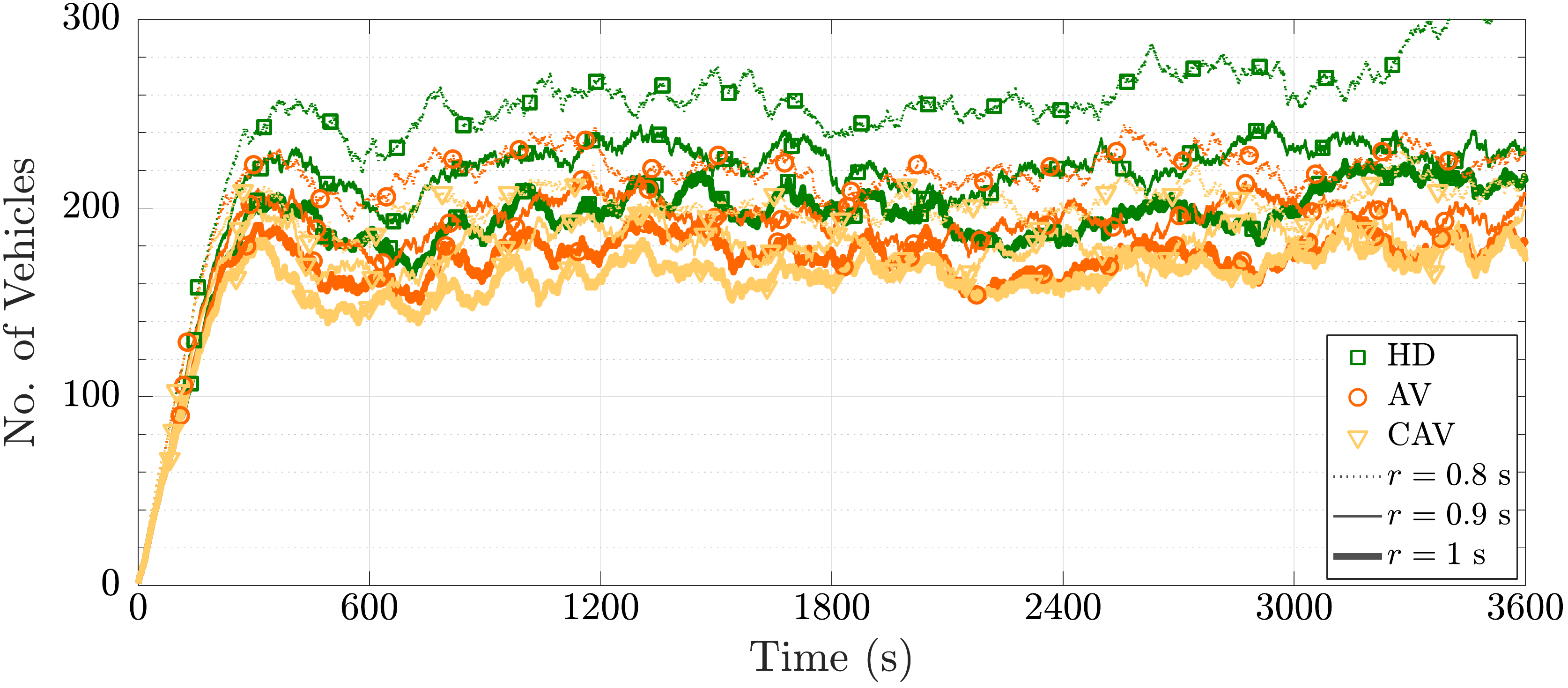}
    \vspace{-5.5mm}
    \caption{Number of vehicles for the uncapped scenario for Manhattan, as a function of time and different insertion rates.}
    \label{fig:uncappedManhattan}
\end{figure}

\begin{figure}[t]     
\centering
\includegraphics[width=1\columnwidth]{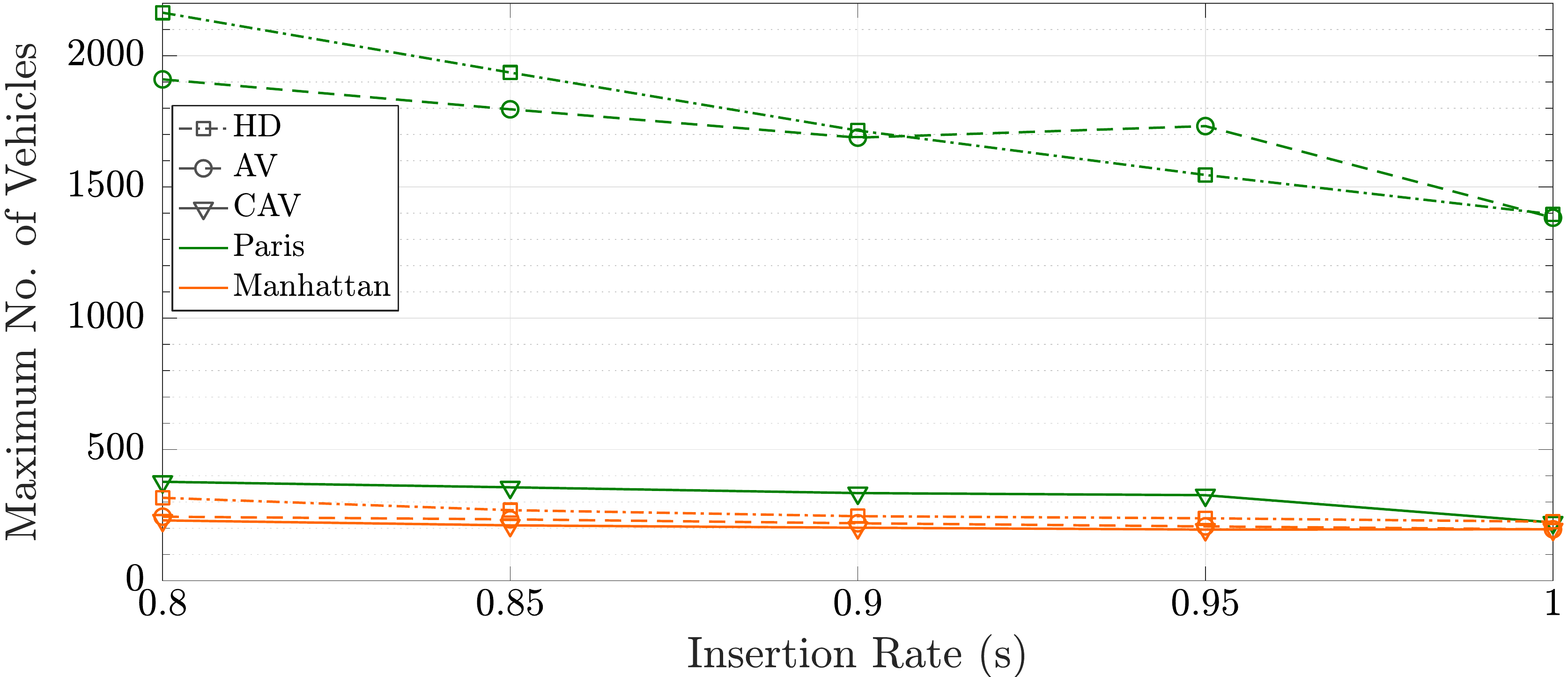}
    \vspace{-5.5mm}
    \caption{Number of vehicles for the uncapped scenario for Paris and Manhattan, as a function of the different insertion rates.}
    \label{fig:uncappedInsertionRate}
\end{figure}

We start with the uncapped scenario (Figs.~\ref{fig:uncappedParis}-\ref{fig:uncappedInsertionRate}).  For Paris, we observe that for both HD vehicles and AVs, as time progresses, the number of vehicles in the scenario is increased. For faster-paced insertion rates, it is shown that the number of AVs is slightly reduced compared to HD vehicles. This is because the enhanced features of an AV (better speed, smaller inter-vehicle distances) alleviate the traffic congestion, and therefore, vehicles arrive at their destination faster. CAVs, however, significantly outperform the other two. This is due to their rerouting capabilities. Having an advanced knowledge about the surrounding environment, they can on-the-fly change their route towards their destination, balancing the traffic on the city. Results for Rome and Berlin followed a very similar behavior with Paris. The above clearly shows the benefits of connectivity, these being the decreased trip time and traffic congestion. 

For London (Fig.~\ref{fig:uncappedLondon}), the difference between CAVs and AVs/HD vehicles is less significant. As described before, London has an increased number of intersections compared to Paris. Vehicles of all types, forced to follow the rules on the road, suffer from decreased average speed and thus increased average journey time. Therefore, the margin of improvement in that scenario is decreased. For Manhattan (Fig.~\ref{fig:uncappedManhattan}), the effect of the layout and the increased number of traffic lights, is even more noticeable, with the gap between the different vehicle types being minimized. However, still, CAVs manage to operate marginally better. Introducing ``virtual'' traffic lights and coordinating the vehicles in smarter ways (for e.g., using a centralized approach), can significantly enhance the results for these grid-like scenarios. Again, the exchange of information between the vehicles and the infrastructure networks will be vital for that.

In Fig.~\ref{fig:uncappedInsertionRate}, the maximum number of vehicles as a function of the insertion rate is shown. As expected, for faster-paced insertion rates, the overall number of vehicles increases. This is the case for all scenarios, except Manhattan, where the city layout and the number of traffic lights, limit the traffic flow significantly. Overall, from the above, we observe that the unique characteristics of CAVs and the ``bird-eye''  knowledge they can acquire, decreases the number of vehicles on the road improving the traffic flow.

\begin{figure}[t]     
\centering
\includegraphics[width=1\columnwidth]{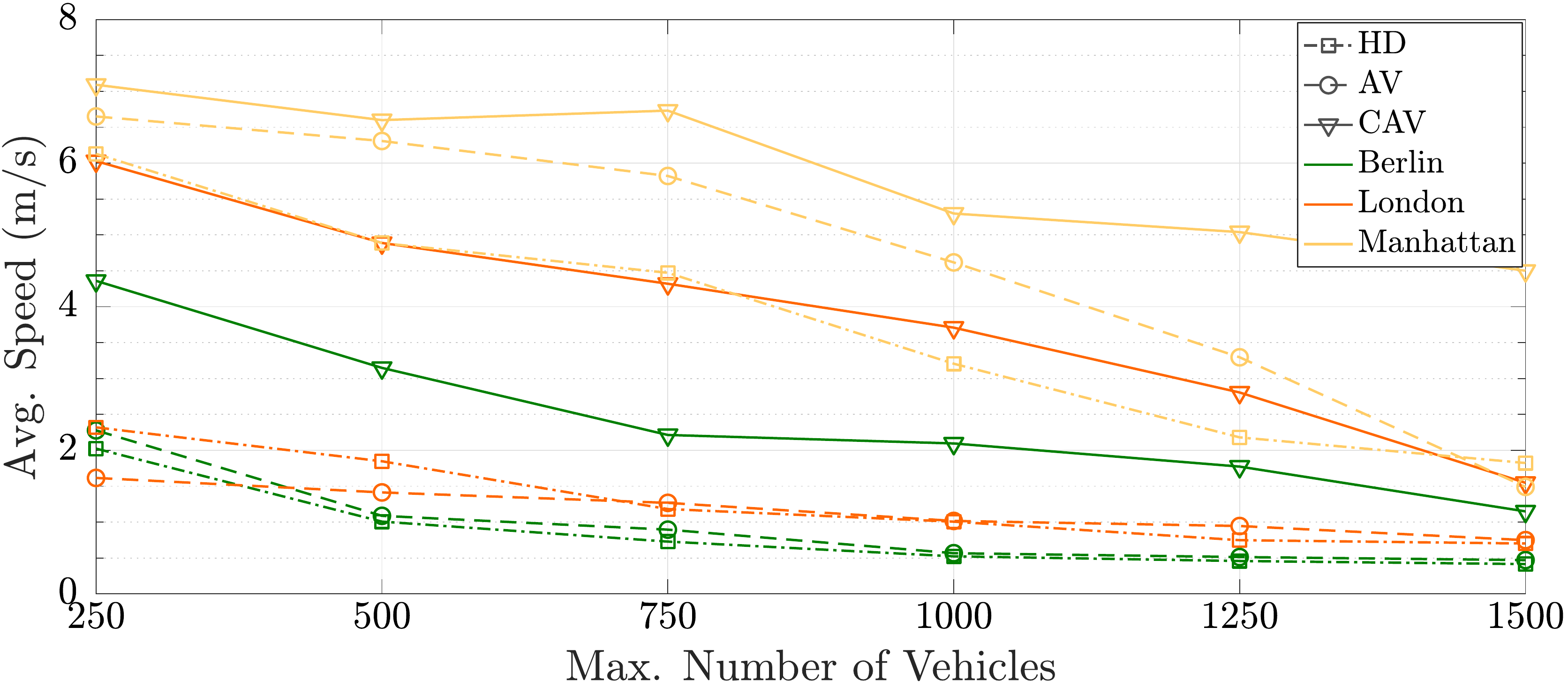}
    \vspace{-5.5mm}
    \caption{Capped scenario for Manhattan, London and Berlin. Average speed as a function of the number of vehicles.}
    \label{fig:cappedAvgSpeed}
\end{figure}

\begin{figure}[t]     
\centering
\includegraphics[width=1\columnwidth]{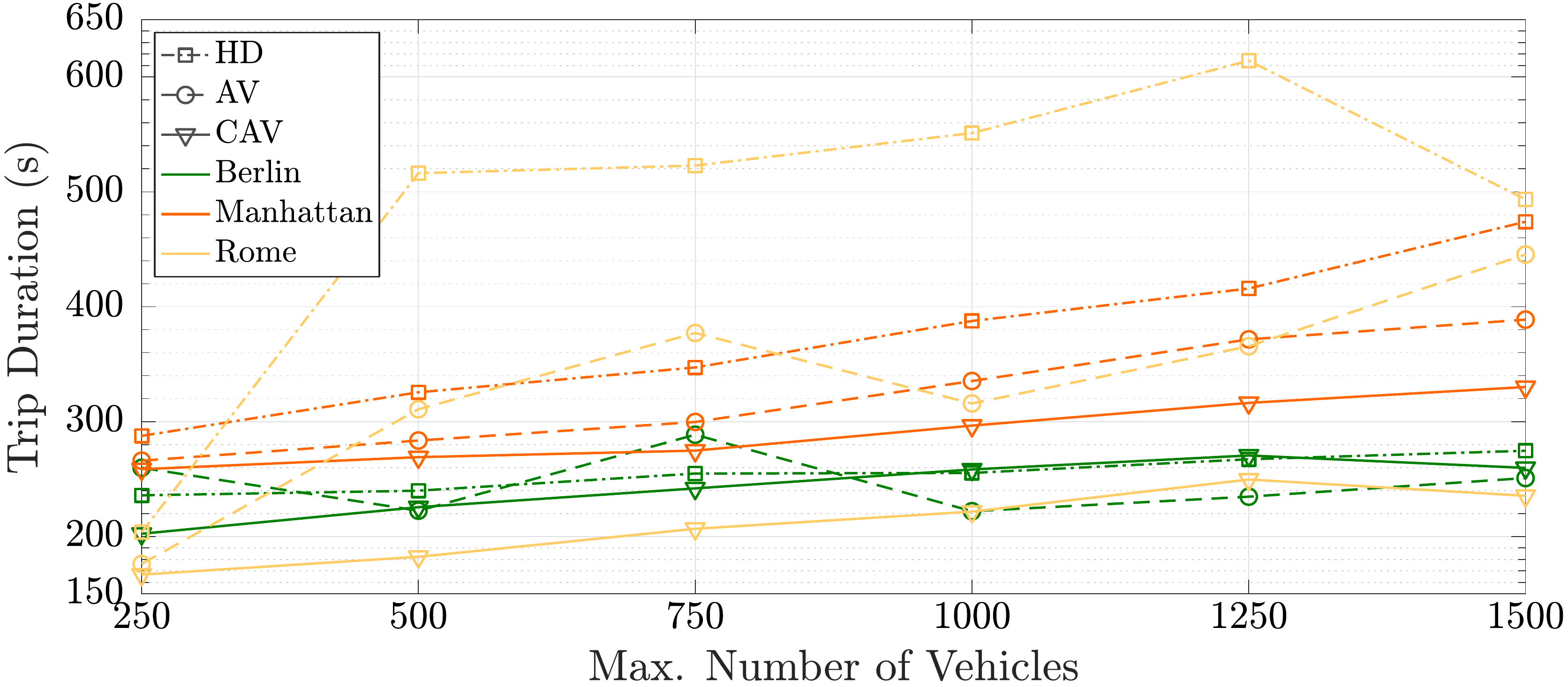}
    \vspace{-5.5mm}
    \caption{Capped scenario for Manhattan, Rome and Berlin. Average trip time as a function of the number of vehicles.}
    \label{fig:cappedTripTime}
\end{figure}

\begin{figure}[t]     
\centering
\includegraphics[width=1\columnwidth]{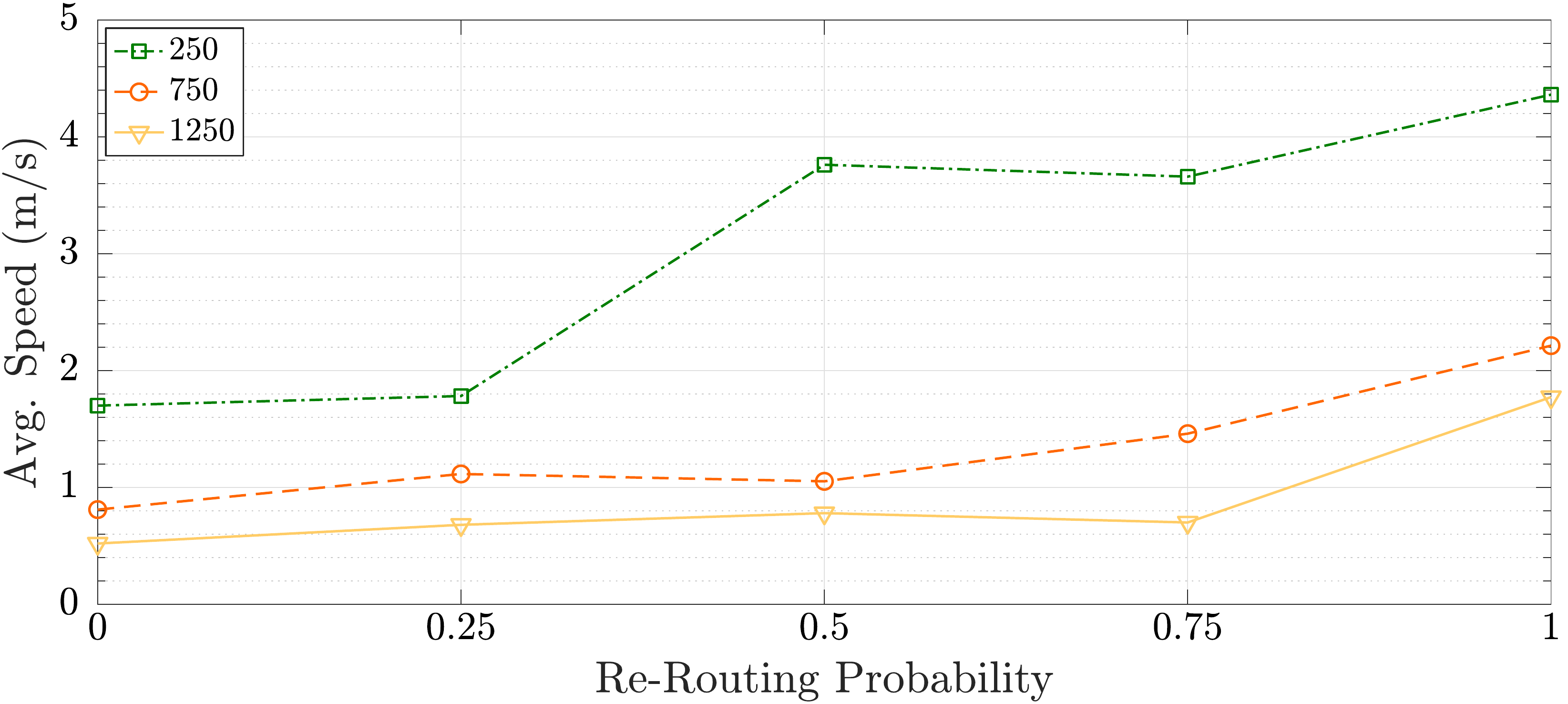}
    \vspace{-5.5mm}
    \caption{Average speed as a function of the re-routing probabilities for Berlin and different maximum number of vehicles.}
    \label{fig:probabilities}
\end{figure}

Figs.~\ref{fig:cappedAvgSpeed}-\ref{fig:probabilities} show the results for the capped scenarios. For all these figures, we consider the fluid bounding box introduced in Sec.~\ref{sub:fluidBox}. Fig.~\ref{fig:cappedAvgSpeed} presents the average speed as a function of the maximum number of vehicles. Starting with London, we observe that HD vehicles and AVs have a very similar performance. On the other hand, the unique characteristic of CAVs significantly improve their average speed. Having advanced knowledge about the traffic jams, they can strategically avoid them, significantly enhancing the overall traffic flow. Results for Rome and Paris are very similar to the London scenario, so they not be presented due to the limited space. 

A similar performance is observed in Berlin as well. Due to the decreased number of traffic lights and road junctions, the improvement margin is decreased compared to London. Again, as before, the average speed gain is decreased as the number of vehicles is increased, and CAVs slightly outperform the other two vehicle types. However, this is not the case for Manhattan. For that scenario, the benefit of CAVs is amplified as the number of vehicles increases. The increased number of traffic lights and junctions cause severe traffic jams in the city. Having the ability to avoid them can significantly reduce the speed loss, and the continuous accelerate/decelerate in traffic.

Fig.~\ref{fig:cappedTripTime} presents the average trip time per vehicle as a function of the vehicle density. As expected, when the average speed increases, the average trip duration time is decreased. This is evident in the results from Manhattan that follow an inversely proportional behavior compared to Fig.~\ref{fig:cappedAvgSpeed}. Moving on to Rome, an initial observation shows that the trip duration follows an upward trend when the number of vehicles increases for all types of vehicles. However, this is not the case for all the different maximum number of vehicles. For e.g., for $1000$ vehicles, the trip time is decreased compared to the $750$ case. This is because of the users equilibrium on the road, also known as Braess's paradox. Braess's paradox describes that for the same number of vehicles, if one more road is added, it is possible to impede the traffic flow. In our case, where the length of the road is constant, we observe a similar behavior when we increase the number of vehicles on the road. Vehicles microscopically interacting on the road (change lanes, give right-of-way at intersections, etc.) and selfishly choose their route can, in some cases, reduce the overall performance. Similar behavior can be observed in Paris and London as well, always having the CAVs to required the minimum average trip time. This phenomenon is even more observable in Berlin, where the road layout and the characteristics of AVs lead them to achieve better performance compared to the CAVs.

Finally, the V2X links formulated are not always reliable, or CAVs may be outside of the coverage road of an RSU. To represent that behavior in our investigation, we reduced the re-routing probability of each CAV per simulation step. The idea is that when a vehicle could not formulate a link or is in outage, it does not have knowledge about the surrounding environment, thus cannot be re-routed on-the-fly. The results for that, and more specifically for the city of Berlin, are presented in Fig.~\ref{fig:probabilities}, showing the average speed as a function of the different re-routing probabilities. As before, the increased traffic congestion always decreases the average speed. However, as traffic congestion increases, reliable connectivity plays a crucial role. For example, when vehicles are capped at $250$, we see that with just $50\%$ of the vehicles capable of formulating communication links, the average speed significantly increases. However, for more congested scenarios, we observe the demand for better connectivity. For example, more than $75\%$ should be able to be connected when we have a capped scenario with $1250$ CAVs, to observe a noteworthy benefit in the average speed, and later in the traffic flows.

\section{Conclusions}\label{sec:conclusions}

In this paper, we investigated the impact of AVs and CAVs on the traffic flows in future cities. Our investigations focused on the demanding urban city layouts of five large-scale road networks. Starting with the data processing pipeline, we presented a way of comparing the diverse city layouts and observed their effect on the different types of vehicles, ensuring fair and informative numerical results. Our performance investigations considered several KPIs related to the traffic flow efficiencies (traffic demand, average speed, average trip time). Our results show that CAVs can significantly benefit the traffic flows, decreasing the average trip times by up to three times under heavy traffic conditions, and hence reducing the traffic congestion. These results complement the existing literature which considered other types of road networks and strengthen the expectations for CAVs and the benefits they will bring. These benefits are potentially even greater, for the case where the trip planning tasks are centralized, which we leave as the future work.  


\section*{Acknowledgment}
This work is funded in part by the Next-Generation Converged Digital Infrastructures (NG-CDI) project, supported by BT Group and EPSRC (EP/R004935/1), in part by the FLOURISH project (Innovate UK, no. 102582), and in part by Toshiba Research Europe Ltd.

\bibliographystyle{IEEEtran}
\bibliography{main.bib,IEEEabrv}

\begin{thebibliography}{10}
\providecommand{\url}[1]{#1}
\csname url@samestyle\endcsname
\providecommand{\newblock}{\relax}
\providecommand{\bibinfo}[2]{#2}
\providecommand{\BIBentrySTDinterwordspacing}{\spaceskip=0pt\relax}
\providecommand{\BIBentryALTinterwordstretchfactor}{4}
\providecommand{\BIBentryALTinterwordspacing}{\spaceskip=\fontdimen2\font plus
\BIBentryALTinterwordstretchfactor\fontdimen3\font minus
  \fontdimen4\font\relax}
\providecommand{\BIBforeignlanguage}[2]{{%
\expandafter\ifx\csname l@#1\endcsname\relax
\typeout{** WARNING: IEEEtran.bst: No hyphenation pattern has been}%
\typeout{** loaded for the language `#1'. Using the pattern for}%
\typeout{** the default language instead.}%
\else
\language=\csname l@#1\endcsname
\fi
#2}}
\providecommand{\BIBdecl}{\relax}
\BIBdecl

\bibitem{MRQ}
``{Global Autonomous Driving Market Outlook},'' Frost \& Sullivan, Tech. Rep.,
  Mar. 2018.

\bibitem{C_ITS}
\BIBentryALTinterwordspacing
``{C-ITS Platform},'' European Commission, Tech. Rep., Jan. 2016. [Online].
  Available:
  \url{http://ec.europa.eu/transport/themes/its/doc/c-its-platform-final-report-january-2016.pdf}
\BIBentrySTDinterwordspacing

\bibitem{NHTSA}
\BIBentryALTinterwordspacing
``{Preliminary Statement of Policy Concerning Automated Vehicles},'' {U.S.
  Department of Transportation, National Highway Traffic Safety Administration
  (NHTSA)}, Tech. Rep., 2013. [Online]. Available:
  \url{http://www.nhtsa.gov/staticfiles/rulemaking/pdf/Automated_Vehicles_Policy.pdf}
\BIBentrySTDinterwordspacing

\bibitem{cooperativeDriving}
F.~{Dressler} \emph{et~al.}, ``{Cooperative Driving and the Tactile
  Internet},'' \emph{Proc. of the IEEE}, vol. 107, no.~2, pp. 436--446, Feb.
  2019.

\bibitem{trafficImplications}
A.~Olia \emph{et~al.}, ``{Traffic Capacity Implications of Automated Vehicles
  Mixed with Regular Vehicles},'' \emph{Journal of Intelligent Transportation
  Systems}, vol.~22, no.~3, pp. 244--262, 2018.

\bibitem{effectReservedLanes}
A.~Talebpour, H.~S. Mahmassani, and A.~Elfar, ``{Investigating the Effects of
  Reserved Lanes for Autonomous Vehicles on Congestion and Travel Time
  Reliability},'' \emph{Transportation Research Record}, vol. 2622, no.~1, pp.
  1--12, 2017.

\bibitem{effectAutonomousVehicles}
\BIBentryALTinterwordspacing
B.~Friedrich, \emph{{The Effect of Autonomous Vehicles on Traffic}}.\hskip 1em
  plus 0.5em minus 0.4em\relax Springer, 2016, pp. 317--334. [Online].
  Available: \url{https://doi.org/10.1007/978-3-662-48847-8_16}
\BIBentrySTDinterwordspacing

\bibitem{impactUrbanNetwork}
Q.~Lu \emph{et~al.}, ``{The Impact of Autonomous Vehicles on Urban Traffic
  Network Capacity: An Experimental Analysis by Microscopic Traffic
  Simulation},'' \emph{Transportation Letters}, vol.~0, no.~0, pp. 1--10, 2019.

\bibitem{reactionTime}
J.~T. Dinges and N.~J. Durisek, ``{Automated Vehicle Disengagement Reaction
  Time Compared to Human Brake Reaction Time in Both Automobile and Motorcycle
  Operation},'' in \emph{WCX SAE World Congress Experience}.\hskip 1em plus
  0.5em minus 0.4em\relax SAE International, Apr. 2019.

\bibitem{sumo}
P.~A. Lopez \emph{et~al.}, ``{Microscopic Traffic Simulation using SUMO},'' in
  \emph{Proc. of ITSC 2018}.\hskip 1em plus 0.5em minus 0.4em\relax IEEE, 2018.

\bibitem{traci4matlab}
A.~F. Acosta, J.~E. Espinosa, and J.~Espinosa, ``{TraCI4Matlab: Enabling the
  Integration of the SUMO Road Traffic Simulator and Matlab{\textregistered}
  Through a Software Re-engineering Process},'' pp. 155--170, 2015.

\bibitem{idmModel}
M.~Treiber, A.~Hennecke, and D.~Helbing, ``{Congested Traffic States in
  Empirical Observations and Microscopic Simulations},'' \emph{Phys. Rev. E},
  vol.~62, pp. 1805--1824, Aug. 2000.

\bibitem{gippsModel}
P.~Gipps, ``{A Behavioural Car-following Model for Computer Simulation},''
  \emph{Transportation Research Part B: Methodological}, vol.~15, no.~2, pp.
  105 -- 111, 1981.

\bibitem{kraussModel}
S.~Krauss, P.~Wagner, and C.~Gawron, ``{Metastable States in a Microscopic
  Model of Traffic Flow},'' \emph{Phys. Rev. E}, vol.~55, pp. 5597--5602, May
  1997.

\bibitem{comparisonModelsMesoscopic}
V.~Kanagaraj \emph{et~al.}, ``{Evaluation of Different Vehicle Following Models
  Under Mixed Traffic Conditions},'' in \emph{Proc. of CTRG 2013}, vol. 104,
  Dec. 2013, pp. 390 -- 401.

\bibitem{comparisonModelsMicroscopic}
L.~Bieker-Walz \emph{et~al.}, ``{Evaluation of Car-following-models at
  Controlled Intersections},'' in \emph{Proc. of ESM 2017}, Oct. 2017.

\bibitem{realisticSUMO}
L.~{Codec{\'{a}}} and J.~{H{\"{a}}rri}, ``{Towards multimodal mobility
  simulation of C-ITS: The Monaco SUMO traffic scenario},'' in \emph{Proc. of
  IEEE VNC 2017}, Nov. 2017, pp. 97--100.

\bibitem{OpenStreetMap}
{OpenStreetMap contributors}, \url{ https://www.openstreetmap.org }, 2017.

\bibitem{speedRelaxation}
M.~Taiebat \emph{et~al.}, ``{A Review on Energy, Environmental, and
  Sustainability Implications of Connected and Automated Vehicles},''
  \emph{Environmental Science \& Technology}, vol.~52, no.~20, pp.
  11\,449--11\,465, 2018.

\bibitem{dynamicSpeedLimits}
\BIBentryALTinterwordspacing
``{Speed and Speed Management 2018},'' European Road Safety Observatory, Tech.
  Rep., 2018. [Online]. Available:
  \url{https://ec.europa.eu/transport/road_safety/sites/roadsafety/files/pdf/ersosynthesis2018-speedspeedmanagement.pdf}
\BIBentrySTDinterwordspacing

\bibitem{studyBehavior}
Y.~{Nishimura} \emph{et~al.}, ``{A Study on Behavior of Autonomous Vehicles
  Cooperating with Manually-Driven Vehicles},'' in \emph{Proc. of IEEE PerCom
  2019}, Mar. 2019.

\bibitem{fluidBoundingBox}
\BIBentryALTinterwordspacing
``{IRF World Road Statistics (WRS) 2012 - 2019},'' International Road
  Federation, Tech. Rep., 2019. [Online]. Available:
  \url{https://www.worldroadstatistics.org/Irf_world_road_statistics_1-disc-2019.php}
\BIBentrySTDinterwordspacing

\end{thebibliography}

\end{document}